# Effect of Grain Orientation and Local Strains on Void Growth and Coalescence in Titanium

Marina Pushkareva[1]; Federico Sket[2]; Javier Segurado[2,3]; Javier Llorca[2,3]; Mohammed Yandouzi[1]; Arnaud Weck[1,2,4,5*]

[1] Department of Mechanical Engineering, University of Ottawa, 150 Louis Pasteur, Ottawa, ON K1N 6N5, Canada

[2] IMDEA Materials Institute, C/Eric Kandel 2, 28906 Getafe, Madrid, Spain

[3] Department of Materials Science, Polytechnic University of Madrid, E. T. S. de Ingenieros de Caminos, 28040, Madrid, Spain

[4] Department of Physics, University of Ottawa, 150 Louis Pasteur, Ottawa, ON K1N 6N5, Canada

[5] Centre for Research in Photonics at the University of Ottawa, 800 King Edward Ave., Ottawa, ON K1N 6N5, Canada

*corresponding author: aweck@uottawa.ca

**Abstract**

Ductile fracture has been extensively studied in metals with weak mechanical anisotropy such as copper and aluminum. The fracture of more anisotropic metals, especially those with a hexagonal crystal structure (e.g. titanium), remains far less understood. This paper investigates the ductile fracture process in commercially pure titanium (CP-Ti) with particular emphasis on the influence of grain orientation and local state of strain on void growth. An experimental approach was developed to directly relate the growth of a void in three dimensions to its underlying grain orientation. Grain orientation was obtained by electron back scattered diffraction on void-containing CP-Ti sheets prior to their diffusion bonding. Changes in void dimensions were measured during in-situ straining within an x-ray tomography system. The strong influence of the embedded grain orientation and that of its neighbors on void growth rate and coalescence has been experimentally quantified. Finite element crystal plasticity simulations that take into account both grain orientation and the local strain state were found to predict the experimental void growth. Grains where basal slip dominates show the largest void growth rates because they are closer to a plane strain condition that favors void growth and coalescence.

**Introduction**

Anisotropy can be found at different scales in metals. At the macroscale, it is caused by the collective grain orientation (texture) that can be found in any material processed, for instance, by rolling, and drawing. At the microscale (grain level), plastic anisotropy is due to the different properties of slip systems, and this anisotropy is particularly strong in hexagonal closed packed structures where prismatic, basal and pyramidal slip can be activated at very different stress levels. This paper focuses on the effect of plastic anisotropy at the crystal level on void growth and coalescence in HCP titanium samples. It is well established that grain orientation can significantly affect mechanical properties of titanium [1–11]. Crystal orientation was shown to affect the friction characteristics of titanium single crystals in vacuum [1] and the fracture toughness of Ti alloys [4]. Viswanathan et al. [7] studied the effect of crystal orientation on the indentation hardness of a Ti alloy using transmission electron microscopy. They noted an increase in hardness near the [0001] indentation axis and attributed the hardening to a high density of ⟨c+a⟩ type geometrically necessary dislocations (GNDs) directly below the indentation. Merson et al. [8] found similar trends, while Britton and al. [9] reported that both indentation modulus and hardness of CP Ti decrease significantly as the indentation axis is inclined further from the c-axis due to the activation of different slip systems. In addition, the anisotropy was more marked in the plastic response. Battaini et al. [5] investigated the effect of sample orientation on the mechanical properties of commercially pure titanium and reported a variation in flow stress of up to 60% for different sample orientations. The higher flow stresses were primarily due to an unfavourable orientation for prism-⟨a⟩ slip and secondly due to a greater proportion of $\{11\bar{2}2\}$ to $\{10\bar{1}2\}$ twinning. While the effect of grain orientation on the strength and hardness of titanium is well documented in the literature, its effect on the ductile fracture of titanium has received much less attention. Lecarme et al. [12] investigated the deformation and fracture of Ti-6Al-4V alloy and observed a large variability in the void growth rates that was related to void shape and crystal orientation effects, as well as to local constraints. Recent work on CP magnesium alloy model materials [13] found that void growth occurs in preferential directions due to the failure at twin and grain boundaries. We also recently showed qualitatively the influence of grain orientation on void growth in CP-Ti using x-ray tomography and crystal plasticity simulations [14].

In addition to the mainly experimental work presented above, numerical approaches have also been used to better understand deformation and fracture in HCP materials. The anisotropy induced by the HCP crystal strcutures have been included in porous metal plasticity models [15-17]. The enhanced growth rates due to crystallographic effects and the activation of certain slip systems was taken into account in ref. [18]. Void growth was also investigated by crystal plasticity, mostly for FCC crystals, [19-22] and more recently for HCP crystals [23,24]. Segurado and Llorca (2010) have shown the effect of void size on their growth in FCC crystals using discrete dislocation dynamics [25].

While the effect of grain orientation on void growth is sometimes inferred in literature and was observed in simulations, there is no direct experimental evidence of the extent to which grain orientation is affecting void growth and this is the main objective of this investigation. To this end, one-to-one comparisons are made in CP-Ti between experimental void growths and crystal plasticity finite element simulations that take into account the initial texture, the underlying grain orientation and the local state of strain.

# 1 Experimental procedure

## 1.1 Sample fabrication

Samples were fabricated in a multi-step process. First, CP-Ti sheets 250 μm thick were micromachined using a femtosecond laser in order to obtain an array of microvoids, 30 μm deep, 30 μm in diameter, and 70 μm apart. One titanium sheet did not contain any void and the second sheet contained 30 μm deep voids. Diffusion bonding of these two sheets resulted in a sample with voids in the bulk. The femtosecond laser system used ensures that clean voids are created with minimum heat affected zone. More details about the machining process is reported in [26]. CP-Ti sheets were then annealed at 870°C during 36 hours under a high vacuum of $10^{-5}$ Pa to promote grain growth up to the point where further annealing would not result in more grain growth. Electron Back Scattered Diffraction (EBSD) was used to obtain the grain orientation in the hole-containing sheet (Figure 1). The EBSD Analysis was conducted using the AZtecHKL software and a Nordlys Detector from Oxford Ins. attached to an EVO-Ma-10 SEM from Zeiss. Diffusion bonding was then carried out to join the sheets together at 870°C during 18 hours under a high vacuum of $10^{-5}$ Pa. This treatment resulted in good bonding between the CP-Ti sheets while minimizing oxidation. After diffusion bonding, the samples were polished and etched to ensure that no subsequent grain growth took place. Fiducial marks were placed on the sample surface using a micro-indenter in order to ensure that the same region was analyzed prior and after annealing. The resulting samples were machined in the shape of dog-bone tensile coupons 2.5 mm wide, 25 mm long, and 0.6 mm thick, with a 0.6 mm wide reduced section in the middle of the gauge length. It should be noted that while the tests were carried out in vacuum, oxygen was still able to reach the sample which resulted in oxygen solid solution strengthening and the suppression of twinning. Indeed, work by Seto et al. [27] has shown that twinning is suppressed when the oxygen content in the sample is larger than 0.11 mass%. The yield strength of the samples in our study is approximately 300 MPa (see Figure 6) which corresponds to an oxygen content > 0.15 mass% and is larger than the amount of oxygen required to suppress twinning. This is the reason why we do not observe twinning experimentally and why twinning was not taken into account in the crystal plasticity simulations carried out in this work.

Samples were tested in tension in-situ in an X-ray tomography system at IMDEA Materials Institute in Madrid, Spain with a custom in-situ tensile rig [28] at a strain rate of $8\times10^{-4}$. Tomograms were collected using a Nanotom 160NF (General Electric-Phoenix) at 90 kV and 170 μA using a tungsten target. For each tomogram, 500 radiographs were acquired with an exposure time of 750 ms. Tomogram voxel size was 1.35 μm. True stress-strain curves were constructed for all samples from the force registered during the tensile test and the smallest cross-sectional area extracted from tomographic reconstructions. To obtain the smallest cross sectional area of the samples in the neck region, the area of each section was calculated automatically along the gage length using the open source image analysis software ImageJ [29]. Void dimensions as a function of true local strain were extracted manually. The true axial (i.e. in the tensile direction) local strain for a given void of interest was obtained from the distance between two adjacent voids in the tensile direction. Similarly, the transverse local strain was determined from the distance between two adjacent voids in the transverse direction (see arrows in Figure 2).

## 2 Experimental results

### 2.1 Microstructural characterization

Annealing CP-Ti at 870°C for a total of 54 hours resulted in an equiaxed structure with large grains (Figure 1(a, b)). EBSD was performed on two samples and the resulting Euler angle map was superimposed on the tomography reconstructions prior to deformation as shown in Figure 1(a, b). Four voids were selected in this paper based on the underlying grain orientation. It was also verified that debonding between Ti sheets did not occur during testing. These voids, together with the grains in which they are located are shown in Figure 1(a, b). Figure 1(c) shows an inverse pole figure identifying the orientation of the four grains. Several EBSD maps were used to generate the texture shown in Figure 1(d). Grain orientations near

⟨0001⟩, i.e. is parallel to the loading direction, and about 45 degrees away were chosen to see the effect of grain orientation on void growth.

**2.2 Void growth**

X-ray tomography allowed the visualization of void growth in three dimensions. The growth of a void in Grain 2 is depicted in Figure 2, where the elongation of the void can be clearly observed. The normalized void radii in both tensile ($R_1/R_0$) and transverse ($R_2/R_0$) directions (where $R_1$ and $R_2$ are the current void radii in the tensile and transverse directions respectively, and $R_0$ is the initial void radius) are presented as a function of the local axial strain in Figure 3. The normalized void growth rate in the tensile direction ($R_1/R_0$) is observed to depend on grain orientation with void growth rate differences between Grain 1 and Grain 2 of about 30%. Grains 1 and 4 grow the fastest while grains 2 and 3 have slower growth rates.

Dimensional changes in the transverse direction ($R_2/R_0$) are small and trends are difficult to establish due to the limited resolution of the X-ray tomography system (voxel size is 1.35 μm). Nevertheless, an initial decrease in the transverse void radius is noticeable for all voids, as expected from a void growing under uniaxial tension prior to strain localization. Afterwards, $R_2$ increases for most voids (except void in grain 2) which suggests that a localization event (e.g. void coalescence) is taking place at the voids.

The evolution of the transverse strain as a function of the axial strain is plotted in Figure 4 for each void in order to characterize the local state of strain around a void. The ratio of transverse strain over axial strain remains constant for a given void but differs significantly among voids within the range 0.54 to 0.88. Voids in Grain 1 and 4 have a ratio of transverse over axial strain closer to 1 which means that these grains are closer to a state of plane strain while voids in Grains 2 and 3 are closer to a state of uniaxial tension.

## 3 Crystal plasticity simulations

### 3.1. Simulation strategy

Crystal plasticity finite element simulations were performed in order to study the effect of grain orientation on void growth in titanium. The modelling strategy consisted in simulating the deformation of a cubic cell containing two regions (Figure 5). A crystal plasticity model [30] is used for the interior region representing a grain containing a spherical void. The shape of the grain corresponds to a tetrakaidecahedron and the ratio between void and grain diameter is (1/4), in agreement with the experimental void diameter and average grain size. The outer region surrounding the grain corresponds to a homogenized elastoplastic material with properties equivalent to the Ti polycrystal and, for simplicity, its behaviour was approximated as an isotropic Von-Mises elastoplastic material. The yield stress and the hardening law for this region were directly taken from the experimental true stress-strain curves.

The crystal plasticity model used for the region representing the grain is an elasto-viscoplastic phenomenological model, as described in [30]. Three sets of slip systems were considered, prismatic, basal and pyramidal <c+a>. The plastic slip rate was obtained using a power law with the same exponent m=0.02 for all the systems considered. Hardening was considered by using a Voce hardening law

$$h(\Gamma) = h_0 + \left(h_0 - h_s + \frac{h_0 h_s \Gamma}{\tau_s}\right) \exp\left(\frac{-h_0 \Gamma}{\tau_s}\right) \qquad [1]$$

where *h* is the hardening modulus, Γ the accumulated shear strain in all slip systems and the parameters τ$_0$, τ$_s$, *h$_0$* and *h$_s$* are fitting parameters defining the CRSS evolution with the accumulated plastic strain. This hardening law allows to represent the asymptotic hardening rate at large strains [31], the parameters τ$_0$ and *h$_0$* describe the initial yield stress and hardening rate in the crystal, respectively, while τ$_s$ and *h$_s$* determine the asymptotic characteristics of the hardening.

The parameters of the model were obtained by combining values from literature and inverse analysis from the stress-strain curve of the two samples considered. The true stress-strain curves were constructed using the force registered during the tensile test and the smallest cross sectional area extracted from tomography reconstructions. The ratios between the initial critical resolver shear stresses (CRSS) of the three systems were taken from experimental data obtained from cantilever microtesting in pure Ti [32] and correspond to $\tau_{basal} = 1.155\,\tau_{prismatic}$ and $\tau_{pyramidal} = 2.618\,\tau_{prismatic}$ where $\tau_{prismatic}$ is an adjustable parameter. Additionally, it was considered that the ratios between the saturated values of CRSS ($\tau_s$) of the three sets were identical to the initial ratios and that $h_0$ and $h_s$ were identical for all the systems. Under these simplifications, the only values to be adjusted by comparison with the macroscopic stress-strain tests were $\tau_{prismatic}, h_0$ and $h_s$. These parameters were obtained by comparison of the experimental stress-strain curves with simulations carried out on polycrystalline cells including the actual texture of the material. The polycrystalline homogenization simulations contained 20x20x20 voxels and 119 grain orientations (more details can be found in [32,33]). The final parameters of the crystal plasticity model are summarized in Table 1. Comparison between the experimental stress strain curves for the two samples considered and the numerical simulations of the polycrystalline models are plotted in Figure 6. It can be observed that the simulations are able to accurately reproduce the macroscopic behaviour by selecting appropriately the three parameters mentioned above.

The simulation of the cell depicted in Figure 5 assumes symmetric boundary conditions and the three components of the overall diagonal deformation tensor were imposed as a relative displacement between the cell sides. Two strain values were then applied to the cells, axial deformation in one direction (y in Figure 5) and transverse deformation in the other (x in Figure 5), while the stress component was zero in the direction perpendicular to the thickness (z in Figure 5). The cell contained 3296 elements.

A simulation is performed for each void characterized in tomography with the grain orientation taken from the actual orientation measured by EBSD. The axial and transverse strains imposed in each simulation correspond to the actual local values of strain (i.e. strain triaxiality) obtained from the analysis of the tomographic images (see Figure 4). Some simulations were also carried out without imposing a transverse displacement to investigate void growth without enforcing the local state of strain found experimentally (i.e. without taking into account the strain triaxiality).

Table 1 : Voce-hardening law parameters for samples S1 and S2.

| Slip system | $\tau_0$ (MPa) | $\tau_s$ (MPa) | $h_0$ (MPa) | $h_s$ (MPa) |
|---|---|---|---|---|
| Prismatic | 86 | 95 | 2500 | 78 |
| Basal | 99 | 109 | 2500 | 78 |
| Pyramidal a+c | 225 | 248 | 2500 | 78 |

### 3.2 Simulations results

Comparison between experimental results and crystal plasticity simulations with and without taking into account the local state of strain (or strain triaxiality, T) are presented in Figure 7. In most cases, the simulations are in better agreement with the experimental results when the local strain state (solid lines in Figure 7) is taken into account. Note that the simulations do not include any localization or damage criterion and thus cannot predict the increase in transverse void radii observed experimentally at high strains.

To better understand the effect of the underlying grain orientation on void growth, the relative activities of the slip systems next to the void were extracted from the simulations for basal, prismatic, and pyramidal slip systems (Figure 8). They show that the relative slip system activity changes depending on the grain orientation and on the local strain applied on the sample. The basal plane is more active for some grain orientations (grain 1 and 4) while prismatic slip dominates deformation in other orientations (grain 2 and 3). Figure 8 also shows the important effect of the local strain state (i.e. strain triaxiality) on slip system activation. For instance, deformation is clearly dominated by prismatic slip (PR in Figure 8(a)) in Grain 1 for simulations that do not take into account the local strain state while basal slip (BA-T) is dominant when the local strain state is included. The same change in major slip system is observed in grain 4 (Figure 8(d)). In the case of grains 2 and 3, prismatic slip is always dominant, regardless of whether the local strains are included or not, but the contributions of the other two slip systems (basal and pyramidal) become more important at higher strains when the local strain state is included. In Figures 7 and 8 some grain orientations were close to each other in order to confirm whether the voids in these grains had the same behaviour in terms of void growth and coalescence.

## 4 Discussion

The effect of grain orientation on void growth was studied indirectly in [14], where a large scatter in void growth was observed experimentally, and it was hypothesized to be due to grain orientation effects. Crystal plasticity finite element simulations showed a similar scatter in void growth results hence supported this hypothesis. In the present investigation, a one-to-one comparison between void growth and grain orientation is possible because the orientation of the grain in which the void is located is known from EBSD measurements. Void growth curves extracted from the tomography experiments (Figure 3) show that there is a scatter in the void growth rate, which is partly explained by differences in grain orientation. This is in agreement with the results in [30] for Ti-6Al-4V alloy, where the large variability in void growth curves was partly attributed to the local crystallographic orientation. Voids in Grain 1 and 4 have similar orientations (as seen in the inverse pole figure in Figure 1(c)) and local strain states (see Figure 4), and thus, are expected to present a similar void growth behavior. The same statement is true for the voids in Grains 2 and 3 that also share similar orientation and local strain state.

Another important effect controlling void growth is the local strain around the void. The local state of strain can change significantly from one void to another due to its underlying grain orientation and that of its neighbouring grains (Figure 4). Thus, simulation results with and without taking into account the local state of strain were performed in order to evaluate the relative influence of grain orientation and local strain state on void growth. The evolution of axial and transverse void dimensions (as given by $R_1/R_0$ and $R_2/R_0$ respectively) as a function of far-field applied strain is shown in Figure 9. These curves indicate that the influence on void growth of the local strain state (i.e. difference between solid lines and dashed lines for a given grain orientation, i.e. given color, in Figure 9) is more important than the grain orientation (i.e. difference between dashed lines in Figure 9). Furthermore, the local state of strain can change the dominating slip system from prismatic to mainly basal in grains 1 and 4 and this change will affect void growth and coalescence. These results emphasise the need to properly describe the local strain state and thus the local environment (e.g. other grains, other voids, etc.) surrounding a particular void if accurate void growth predictions are to be made. It should also be noted that grain orientation and local state of strain are coupled, e.g. a grain in a given orientation may deform more (or less) which in turns will affect the local state of strain.

The experimental results in Figure 3 indicate that the voids first elongate in the tensile direction with a decrease in the transverse void radius. Voids then grow towards each other in the transverse direction when interactions between voids start and coalescence occurs (which is shown by an increase in transverse void radius). Although the data in the transverse direction are noisier, they still show that the increase in transverse void radius takes place much earlier for voids in grains 1 and 4 (at a strain of about 0.055 in both cases) than for grains 2 (no increase) and grain 3 (at a strain of about 0.12). These results suggest that coalescence takes place earlier in grains 1 and 4 compared to grains 2 and 3. It is worth mentioning that grains 1 and 4, where basal slip dominates, also have the highest void growth rates (Figure 3) and the highest ratio of transverse over axial strains suggesting that these two grains are closer to a plane strain condition, which favors strain localization by shear band formation and thus coalescence.

**Conclusion**

- This paper presents a novel methodology to study the effect of grain orientation and local state of strain on void growth. New model materials were fabricated by laser micromachining and diffusion bonding. X-ray tomography was used to visualize void growth and to obtain the local strain state in three-dimensions during in situ tensile tests while the grain orientation for each void was obtained by electron back-scattered diffraction (EBSD). This allowed, for the first time, to directly couple void growth with grain orientation and local strain state, in three dimensions.

- The influence of the local strain state around a void has a larger impact on void growth than the orientation of the grain in which the void is located (for the grain orientations studied here). In other words, the influence of first neighbor grains on the growth rate is higher that the local grain orientation due to the constraint they impose in the plastic deformation.

- Crystal plasticity simulations that take into account the particular grain orientation and the local strain state around the voids were able to predict void growth rates accurately.

- For some grain orientations, the local strain state has a significant effect on the dominant slip system, which in turn affects void growth and coalescence.

- The highest void growth rates were found in grains oriented for basal slip because the deformation of these grains was closer to a plane strain condition, which favors shear band formation, void growth, and coalescence.

- The results presented here highlight the importance of taking into account both grain orientation and the local strain state in order to obtain accurate void growth and coalescence predictions.

This study and its findings are important not only for CP Ti used in aerospace and biomedical industries but also for other materials exhibiting properties strongly depend on grain orientation and local strain state.

**Acknowledgements**

AW acknowledges support from the Natural Sciences and Engineering Research Council of Canada (NSERC) as well as the International Fellowship MICROFRAC (Visualization and modeling of fracture at the microscale) supported by the European Union, H2020 program, Marie Skłodowska-Curie Actions (contract 659575). This investigation was supported by the European Research Council (ERC) under the European Union's Horizon 2020 research and innovation program (Advanced Grant VIRMETAL, grant agreement No. 669141). JS also acknowledges the financial support received from the Spanish Ministry of Economy and Competitiveness under grant DPI2015-67667-C3-2-R.

**Data availability**

The raw/processed data required to reproduce these findings cannot be shared at this time due to technical or time limitations.

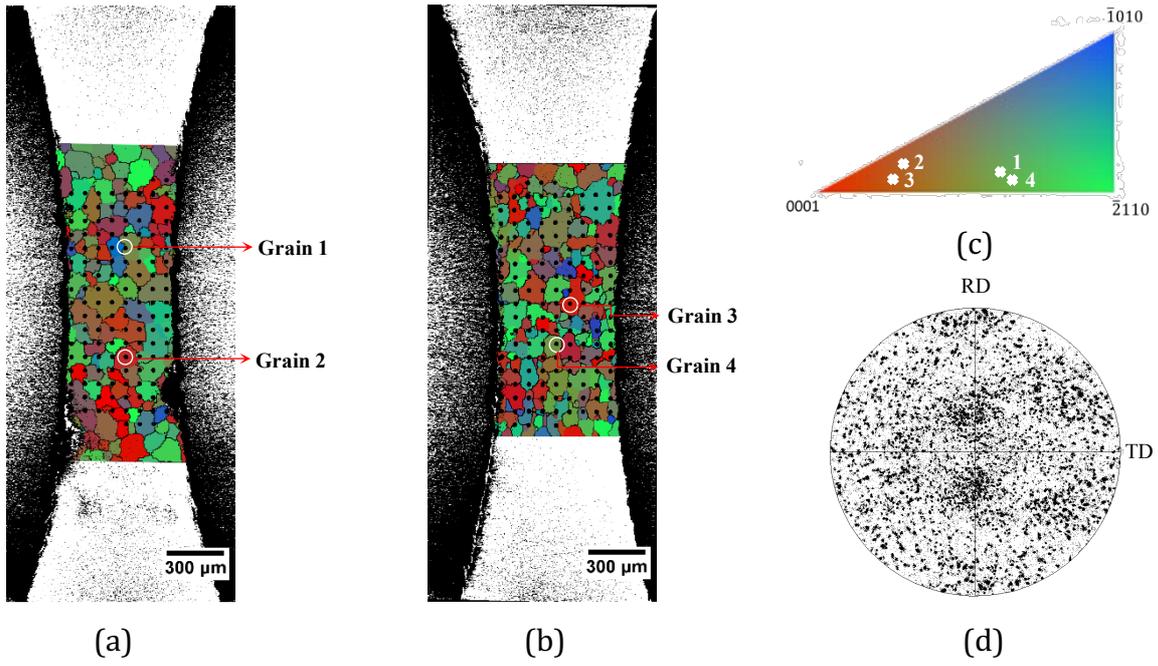

*Figure 1: Initial tomography reconstruction superimposed with its corresponding Euler angle EBSD map for: (a) sample S1, (b) sample S2. (c) Inverse pole figures identifying the selected grain orientations for samples S1, S2. (d) Pole figure based on ESBD maps of samples S1 and S2. Tensile direction is RD or 〈0001〉. RD is rolling direction, TD is transverse direction.*

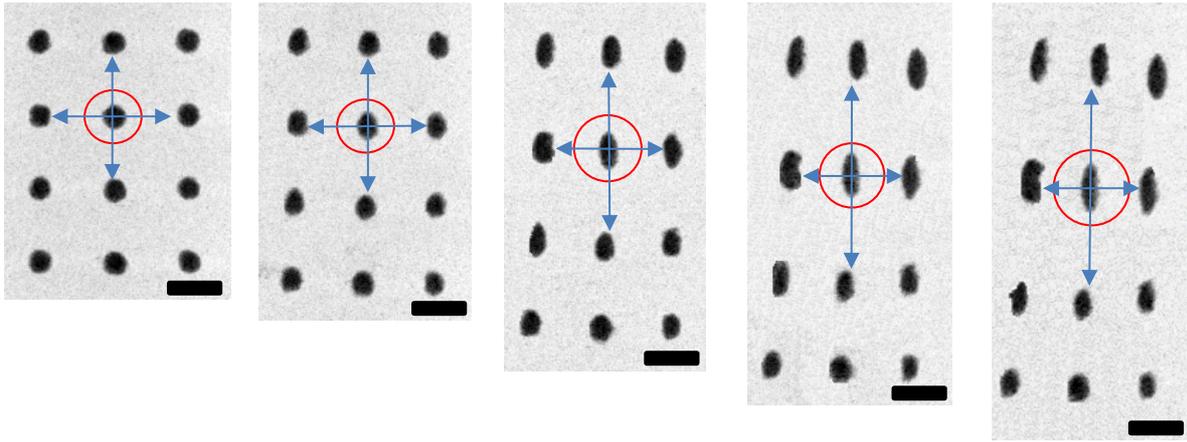

(a) 0 / 0    (b) 0.10 / 0.06    (c) 0.26 / 0.13    (d) 0.41 / 0.23    (e) 0.47 / 0.24

*Figure 2: Tomography reconstructions showing the growth of a void in grain 4. The circles indicate the voids of interest. The numbers indicate the local axial and transverse strains calculated using the distance between adjacent voids (arrows). Tensile direction is vertical. Scale bar is 50 μm.*

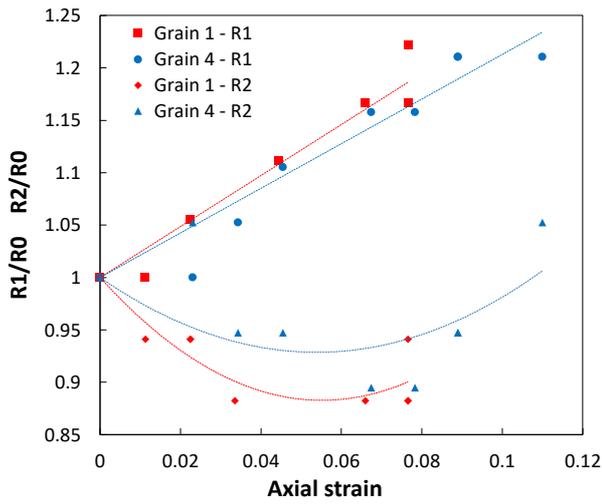 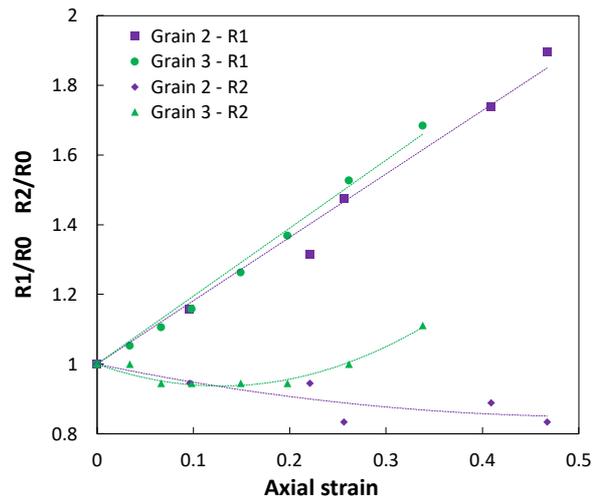

(a)                  (b)

*Figure 3: Experimental result of void growth as a function of local strain in the loading direction for voids in grains (a) 1 and 4, and (b) 2 and 3. R1 and R2 are the current void radii in the tensile and transverse directions respectively, and R0 is the initial void radius. The experimental data are fitted with straight dotted lines for R1 and with dotted polynomials for R2 for clarity.*

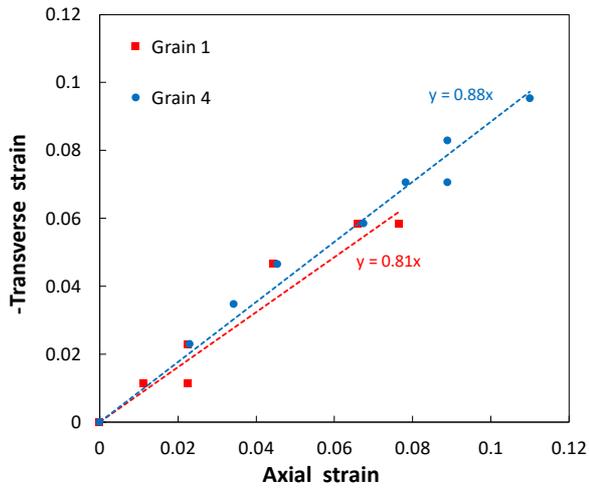 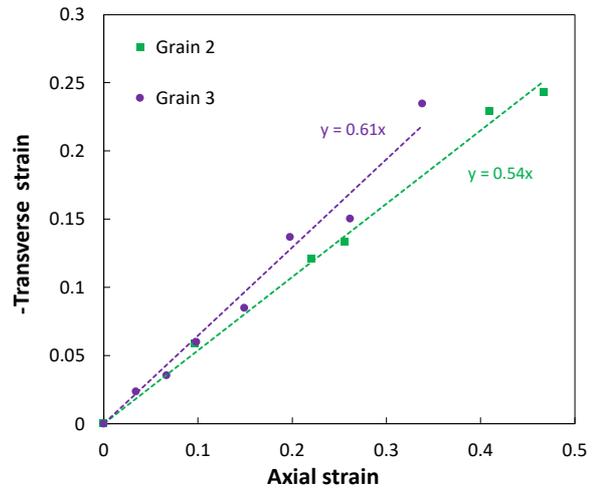

(a)                  (b)

*Figure 4: Transverse strain as a function of axial strain in different voids. The experimental data (dots) are fitted with straight dotted lines for clarity.*

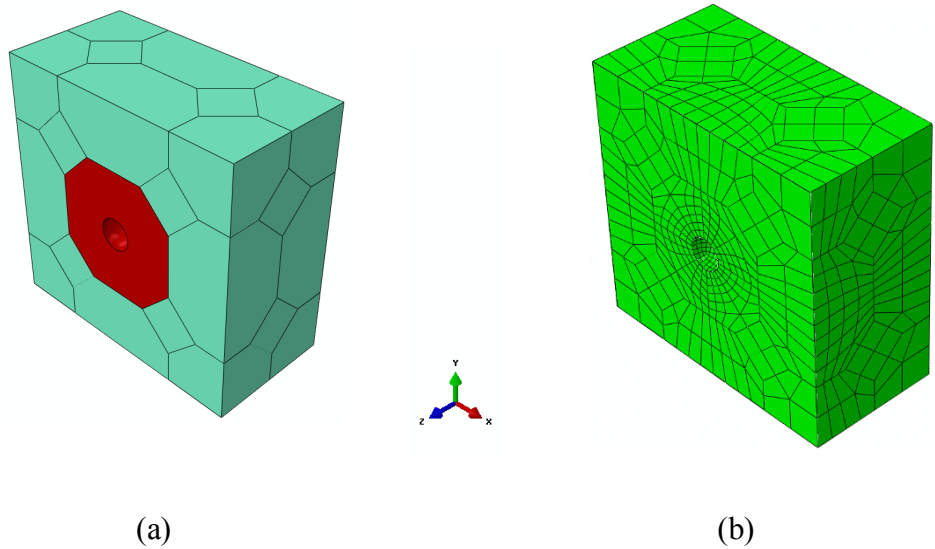

(a)                        (b)

*Figure 5: 3D model with one void at the center of one grain. Tensile direction coincides with the y axis and the model is symmetric with respect the XY plane. (a) Red color indicates crystal plasticity; green color indicates macroscopic properties of titanium, (b) Finite element discretization.*

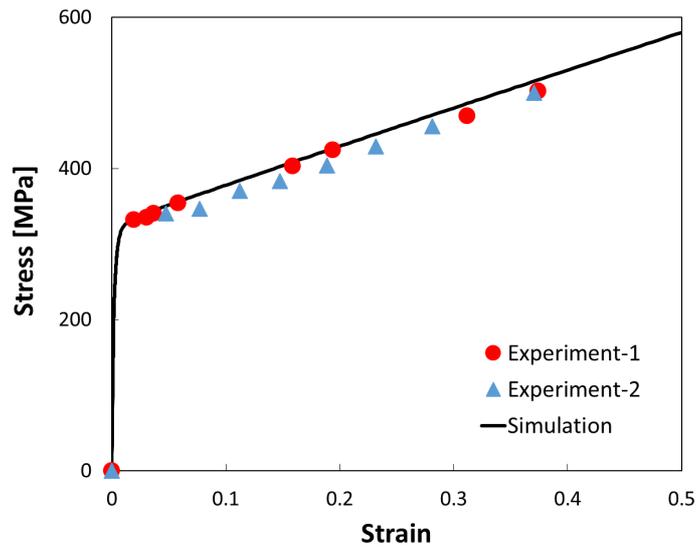

*Figure 6: Tensile stress-strain curves: experimental results and numerical simulation.*

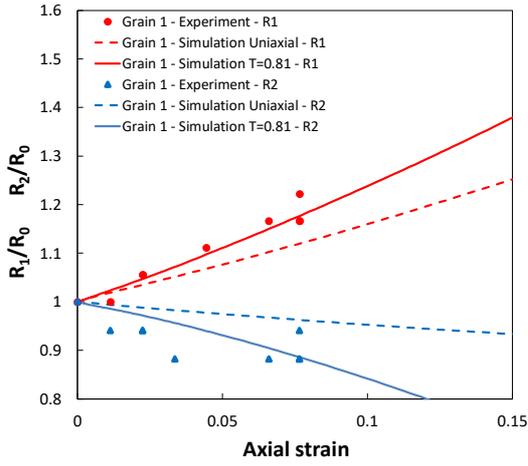

(a)

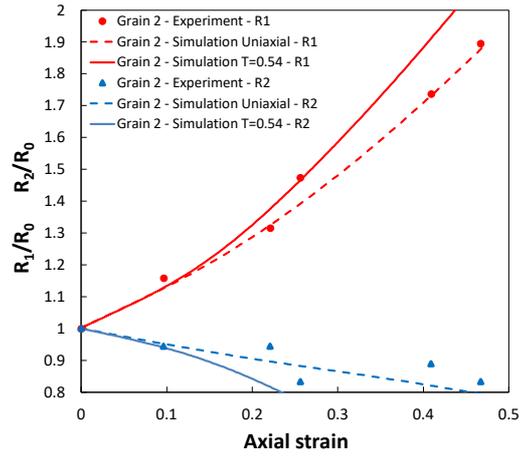

(b)

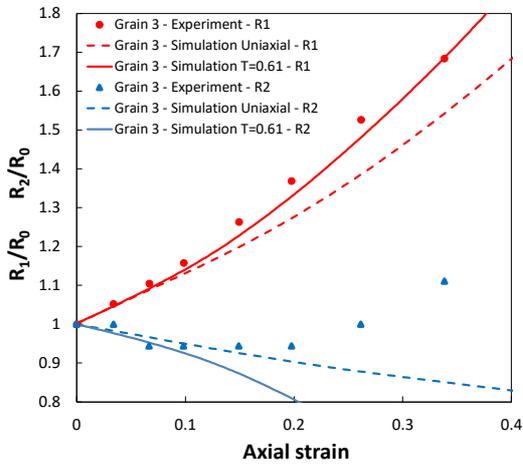

(c)

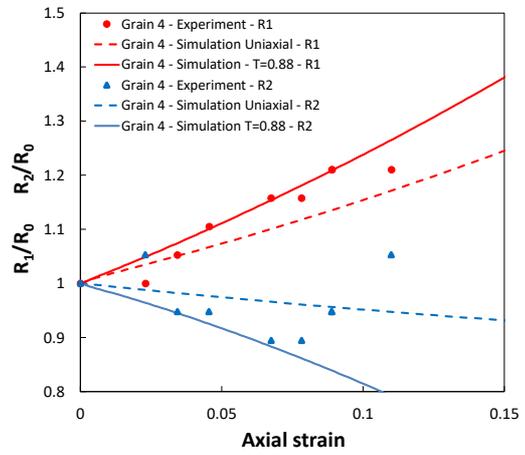

(d)

*Figure 7: Crystal plasticity (solid and dashed lines) and experimental (dots) void growth results as a function of local axial strain in: (a) Grain 1, (b) Grain 2, (c) Grain 3, and (d) Grain 4. The broken lines correspond to simulations carried out without taking into account the local strain state ('uniaxial') while the solid lines included the local strain state or strain triaxiality 'T'.*

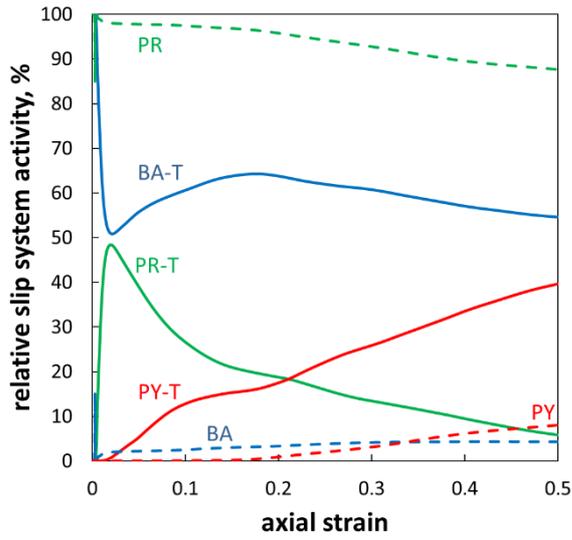
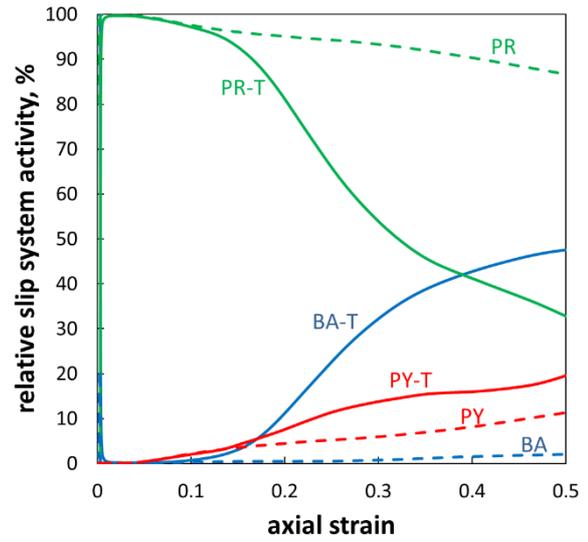

(a)  (b)

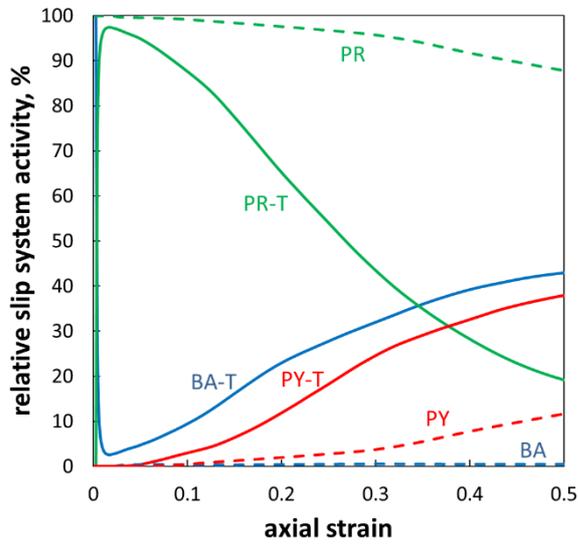
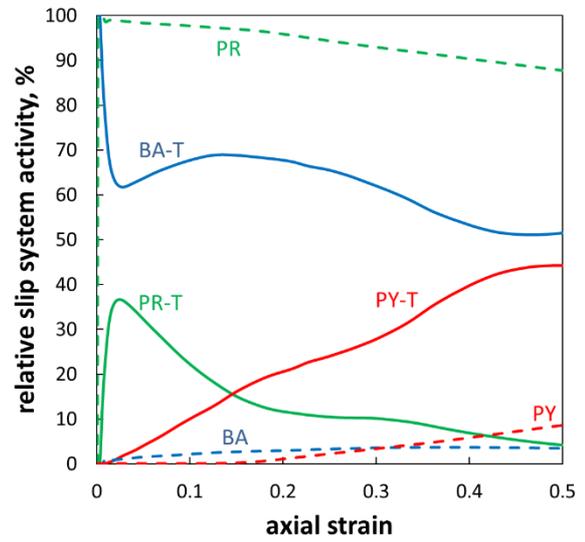

(c)  (d)

*Figure 8: Relative slip system activity in grains: (a) grain 1, (b) grain 2, (c) grain 3 and (d) grain 4. The dashed lines are for simulations that do not take into account the local strain state (or strain triaxiality) while the solid lines include this effect. PR=Prismatic, PR-T=Prismatic with strain triaxiality, BA=Basal, BA-T=Basal with strain triaxiality, PY=Pyramidal, PY-T=Pyramidal with strain triaxiality.*

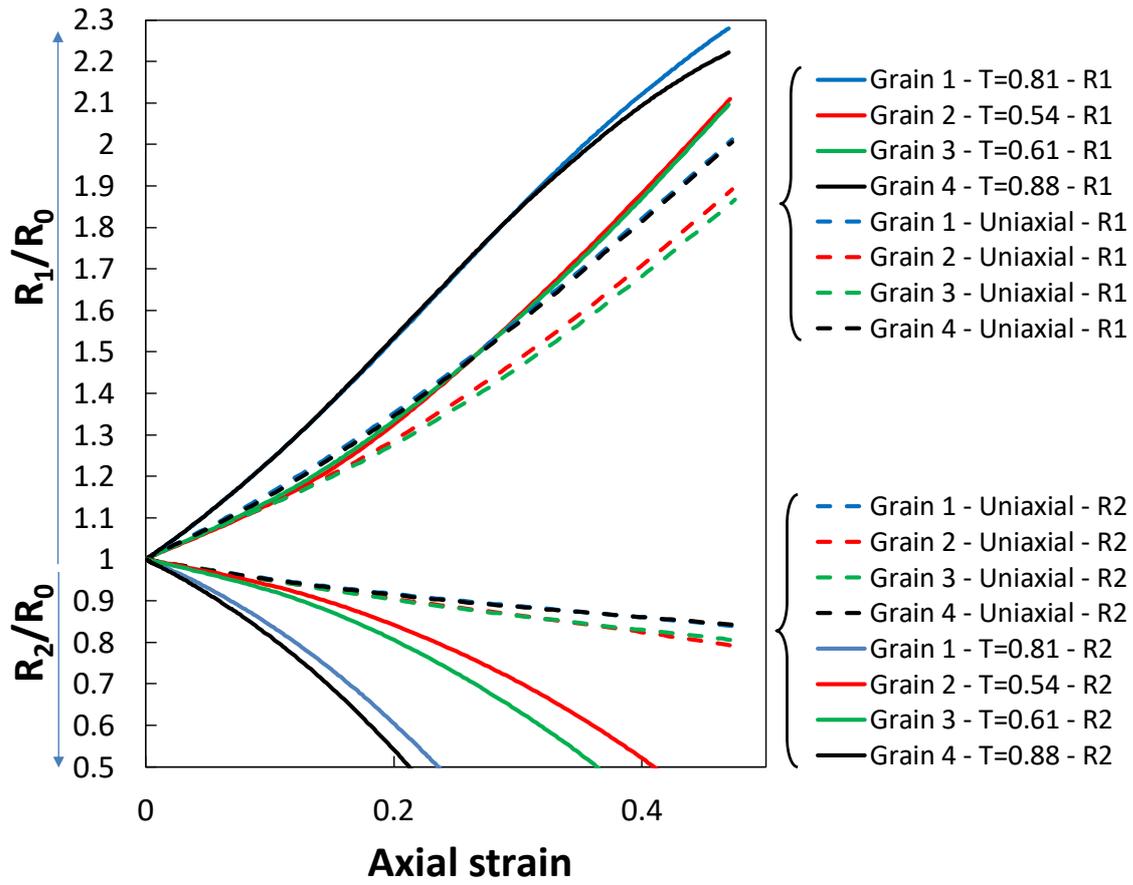

*Figure 9: Crystal plasticity void growth evolution as a function of local strain for different grain orientations, with and without taking into account the local state of strain.*